\documentclass[12pt]{article}
\usepackage{amsmath}
\usepackage{epsfig}
\begin{document}
\title{QCD sum rules for the nucleon and the non-locality of quark condensates
}
\author{M. G. Ryskin, V. A. Sadovnikova\\
{\em National Research Center "Kurchatov Institute"}\\
{\em B. P. Konstantinov Petersburg Nuclear Physics Institute}\\
{\em Gatchina, St. Petersburg}}
\date{\today}
\maketitle
\begin{abstract}
We emphasize the role of the $x$ dependence of the quark condensate $\langle |\bar q(0)q(x)|\rangle$ originated by the zero fermion modes in the instanton vacuum. The effect is not negligible since the inverse size of the instanton $1/\rho\simeq 600$~MeV is comparable with the value of Borel mass $M\sim 1$ GeV.  Using as an example the sum rules for the nucleon we show  that this "non-locality" may strongly affected the conventional results obtained under the $\langle |\bar q(0)q(x)|\rangle=\langle |\bar q(0)q(0)|\rangle$ approximation.
\end{abstract}

\section{Introduction}
The idea of the sum rules (SR) approach is to express the characteristics of the observed hadrons in terms of the
vacuum expectation values of the QCD operators, often referred to as the condensates. It was suggested in \cite{1}
for calculation of the characteristics of mesons. Later it was used for the nucleons \cite{2}.
It succeeded in calculation of the nucleon mass, anomalous magnetic moment, axial coupling constant, etc. \cite{3}.

The QCD sum rules  approach is based on the dispersion relation for the
function describing the propagation of the system which carries the quantum
numbers of the hadron. This function is usually referred to as the "polarization operator"  $\Pi(q)$,
with $q$ the four-momentum of the system. The dispersion relation in which we do not take care of the subtractions reads
\begin{equation}
\Pi(q^2)=\frac1\pi\int dk^2\frac{\mbox{Im}\,\Pi(k^2)}{k^2-q^2}.
\label{0}
\end{equation}

 Thanks to the asymptotic freedom of  QCD the polarization operator, $\Pi(q^2)$, on the left hand side (LHS) can be calculated at high {\em negative} momentum $q^2$.
The large distances contribution is included within the framework of the Operator Product Expansion (OPE) in terms of the vacuum expectation values of some local operators like $\langle |\bar q(0)q(0)|\rangle,\ \langle |\frac{\alpha_s}{\pi}G^{a \mu\nu}(0)G^a_{\mu\nu}(0)|\rangle,\ ...$, so-called "condensates". As a rule the expectation values of these operators, $\langle |O_n|\rangle$, are of the order of $(250\,{\rm MeV})^n$, where $n$ is the dimension of the corresponding operator. So at a large $-q^2$ the expansion of the type $\sum \langle |O_n|\rangle/q^n$ is well convergent.

On the other hand
 the same quantity may be calculated
as the dispersion integral (\ref{0}) over the {\em positive} $k^2$
 where the discontinuity,  ${\rm Im}\,\Pi$, is given by the lowest pole (that is the contribution of our hadron
of interest) and some heavier (continuum) states assuming that the "continuum"  contribution is described by the same expressions as that used on the LHS. In other words something like the "local parton-hadron duality" is assumed.

The parameters which phenomenologically describe the ${\rm Im}\,\Pi(k^2)$ on the RHS of (\ref{0}) are tuned in such a way to
reproduce (as well as possible) the LHS calculated within the perturbative QCD and OPE approach. Of course, in a relatively simple parametrization this in not possible in the whole $q^2$. It becomes possible in some limited interval, say, $q^2_{left}< q^2 < q^2_{right}$, called the duality interval. The corresponding values of $|q^2|$ in this duality interval should not be too small in order to justify pQCD and to provide a good OPE convergency on the LHS. They should not be too large -  not too loose the sensitivity to the parameters of the pole contribution in the RHS.

The polarization operator can be written as
\begin{equation}
\Pi(q^2)=i\int d^4xe^{i(q\cdot x)} \langle 0|T[j(x)\bar j(0)]|0 \rangle,
\label{6}
\end{equation}
where $j(x)$ is the local operator with the hadron quantum numbers,
often referred to as ``current". It is a composition of the quark operators.

To suppress further the higher mass (continuum) contribution, the Borel transform is used, which is equivalent to  inclusion of an additional factor e$^{-k^2/M^2}$ in the integrand
in the dispersion integral (\ref{0}). The Borel transform converts the functions of $q^2$ to the functions
of the Borel mass $M^2$. Note also that the Borel transform removes the divergent contributions caused by behavior of the integrand in the integral on the RHS of Eq.~(\ref{6}) at the lower limit. The hope is that there
is an interval of the values of $M^2$ where the two sides of the SR have a good
overlap, approximating also the true functions.

To calculate the polarization operator defined by Eq.~(\ref{6}) we must clarify the form of the current $j(x)$. Following~\cite{2,9} we use for the proton
\begin{equation}
j(x)=(u^T_a(x)C\gamma_{\mu}u_b(x))\gamma_5 \gamma^{\mu}d_c(x) \varepsilon^{abc}.
\label{8}
\end{equation}
The strong point of this choice is that it makes  the domination of the lowest pole contribution in the RHS of (\ref{0}) more pronounced.\\

Now let us come back to the convergency of the OPE.
The polarization operator (\ref{6}) includes the fields measured at the coordinates $x$ and 0, like $\langle |\bar q(0)q(x)|\rangle$. However
 the  contribution of nonlocal expectation value  $\langle |\bar q(0)q(x)|\rangle$
is usually replaced by the value  $\langle |\bar q(0)q(0)|\rangle$.

Strictly speaking, the original operator is not local and may have some $x$ dependence. If the $x$ dependence is rather flat we may neglect it or to add few first terms of the $x^2$ expansion which looks symbolically\footnote{To provide the gauge invariance we have to use actually the covariant ("long") derivative $D_\mu$.} like $\langle |\bar q(0)q(x)|\rangle=\langle |\bar q(0)q(0)|\rangle+x^2 d\langle |\bar q(0)q(x)|\rangle/dx^2|_{x=0}+...$; the characteristic value of $x^2$ here is driven by the exponential factor in (\ref{6}), that is $x^2\sim 1/q^2$. However when the original "condensate" varies at the scale close to the Borel mass $M^2$ we have to take care about this ($x$-dependence) more seriously~\footnote{
Recall that the nonlocal expectation values
$\langle  0|\bar q(0)q(x)|0\rangle$ is not gauge invariant. To deal with the gauge invariant quantities it should be supplemented by the $P$-exponent from $x$ to 0.
However the current $j$ in (\ref{6}) is colorless and the whole $P$-exponent which accounts for all three quark lines is equal to 1. Therefore below we shall omit these $P$-exponents factors. Moreover, since actually
we shall deal just with zero mode of the quark wave function in the instanton gluon field, already averaged over the colour indexes, it is sufficient to include the effect of the $\langle  0|\bar q(0)q(x)|\rangle$ contribution as some dynamical mass in the original quark propagator.}.

In terms of OPE the value of the corresponding vacuum expectation of operator
$\langle|\bar qD^2q|\rangle$ was evaluated to be $r_q = \langle|\bar qD^2q|\rangle/\langle|\bar qq|\rangle = 0.4\pm 0.1$ GeV$^2$~\cite{BI} (here $D$ is the covariant derivative). However a three time larger value was obtained in the instanton-liquid model~\cite{Si}. That is actually the parameter of expansion $r_q/M^2\sim 1$ is rather large and it may be not sufficient to account just for a few first terms. The structure of the condensates should be considered in more details~\footnote{For the case of wave function of a pion  the role of nonlocal condensates was discussed in~\cite{MR}.}.

What is the origin of the $\langle |\bar q(0)q(x)|\rangle$ condensate? Usually it is assumed that the QCD vacuum is filled by instantons and the vacuum expectations $\langle |\frac{\alpha_s}{\pi}\,G^{a\,\mu\nu}G^a_{\mu\nu}|\rangle$ and $\langle |\bar qq|\rangle$ are caused by the instanton gluon field and the quark zero mode in this instanton field (see, e.g., \cite{5,7,8}). Therefore we consider the instanton vacuum model. The problem is that the typical size of the instanton $1/\rho\sim 600$~MeV is not much larger than the inverse Borel mass $M\sim 1$\,GeV. This means that assuming that the condensates are mainly originated by the instantons we can not neglect the $x$-dependence of the corresponding operators.
 The new dimensionless parameter on the LHS of (\ref{0}) is $\rho^2\,q^2\sim1$ or in terms of the Borel mass  the parameter $\rho^2M^2$ is of the order $\rho^2M^2\sim 1$ at the region of the suitable values of $M^2$. Thus we have to account for the fact that the "condensates" are not homogeneous and its expectation values can not be replaced by the local quantities taken at $x=0$.\\

In the present paper we study the role of such a "non-locality" in SR for nucleon based on the instanton vacuum model where the $\bar qq$ condensate is formed by the quark zero mode wave functions around the instanton. In Sec.~2 we recall the main details of QCD sum rules for the nucleon. In Sec.~3 we consider a simplified model
where the $x$-dependence of the instanton induced condensates is included explicitly. The results are presented in  Tables 1,2 and are discussed in Subsec.~3.3. Some comments about the previous attempts to account for the instanton effects in QCD sum rules for the nucleon are placed in Sec.~4. We conclude in Sec.~5. Clearly, the structure of the QCD sum rules, that is the relative contributions of different terms, are noticeably affected by accounting for the condensates non-locality.  However the QCD sum rules still may reproduce the value of  nucleon mass observed experimentally.

\section{QCD sum rules for nucleon without the instantons}
In the case of nucleon (we consider the proton) the polarization operator takes the form
\begin{equation}
\Pi(q)=\hat q\Pi^q(q^2)+I\Pi^I(q^2),
\label{1}
\end{equation}
with $q$ the four-momentum of the system, $\hat q=q_{\mu}\gamma^{\mu}$, $I$ is the unit matrix. Now we have the dispersion relations for each component.
 \begin{equation}
\Pi^i(q^2)=\frac1\pi\int dk^2\frac{\mbox{ Im}\,\Pi^i(k^2)}{k^2-q^2}\,;
 \quad i=q,I
\label{2}
\end{equation}
(we do not take care of the subtractions).

The left hand side  of Eq.~(\ref{2}) is calculated as the OPE series. The imaginary part on the right hand side  describes the physical states with the proton quantum numbers. That is in  ${\rm Im}\,\Pi(k^2)$ we have to include the proton pole and  the cuts corresponding to the systems containing the nucleon and pions, etc.
Usually the RHS of Eq.~(\ref{2}) is approximated by the "pole+continuum" model
\cite{1,2} in which the lowest lying pole is written down exactly,
while the higher states are described by the same perturbative contribution as that on the LHS but only starting from some continuum threshold $W^2$.
Thus Eqs.~(\ref{2}) take the form
\begin{equation}
\Pi^{i~OPE}(q^2)=\frac{\lambda_N^2\xi^i}{m^2-q^2}+\frac1{2\pi i}
\int\limits_{W^2}^\infty dk^2\frac{\Delta\Pi^{i~OPE}(k^2)}{k^2-q^2};
\quad i=q,I.  \label{3}
\end{equation}
Here $\xi^q=1$,\ $\xi^I=m$. The upper index OPE means that several lowest OPE terms are included.
Note that the ``pole+continuum"  presentation of the RHS makes sense only if its
first term, treated exactly, is larger than the second term, which
approximates the higher states.  The position of the lowest pole $m$,
its residue $\lambda_N^2$ and the continuum threshold $W^2$ are the
unknowns of  Eqs.~(\ref{3}). These parameters are fitted in such a way to provide the (as possible) "exact" equality between the LHS and the RHS of (\ref{3}) in a limited region of $q^2$.

 The corresponding fitting procedure called the "solution" of (\ref{3}) and
the limited region of $q^2$  -- the duality interval. One usually applies the Borel transform which converts the functions of $q^2$ to the functions
of the Borel mass $M^2$. An important assumption is that there
is an interval of the values of $M^2$ where the two sides of the SR have a good
overlap, approximating also the true functions. This interval is in the region of $1\,{\rm GeV}^2$. For a lower $M^2$ we loose the accuracy of calculation on the LHS; the convergency of the OPE  becomes worse and the higher $\alpha_s$ order QCD corrections increase. For a larger $M^2$ the sensitivity of the RHS to the proton pole contribution falls down.  Thus actually
one tries to expand the OPE  from the high momentum region to the region of $1\,{\rm GeV}^2$.

To calculate the LHS the polarization operator (\ref{6}),  the Ioffe current (\ref{8}) is used. Then the LHS of Eq.~(\ref{3}) can be written as
\begin{equation}
\Pi^{q~OPE}(q^2)=\sum_{n=0}A_n(q^2); \quad \Pi^{I~OPE}(q^2)=\sum_{n=3} B_n(q^2)
\label{3a}
\end{equation}
where the lower index $n$ is the dimension of the corresponding QCD condensate ($A_0$ stands for the three-quark loop).
The most important terms for $n\leq 6$ were obtained earlier \cite{2,3}
\begin{equation}
A_0=\frac{-Q^4\ln{Q^2}}{64\pi^4}; \quad A_4=\frac{-b\ln{Q^2}}{128\pi^4}; \quad A_6=\frac{1}{24\pi^4}\frac{a^2}{Q^2};
\quad B_3=\frac{aQ^2\ln{Q^2}}{16\pi^4},
\label{3c}
\end{equation}
with $Q^2=-q^2 >  0$, while $a$ and $b$ are the scalar and gluon condensates multiplied by certain numerical factors
\begin{equation}
a=-(2\pi)^2\langle  0|\bar q q|0\rangle  ; \quad b=(2\pi)^2\langle  0|\frac{\alpha_s}{\pi}\,G^{a\mu \nu}G^a_{\mu \nu}|0\rangle .
\label{3d}
\end{equation}
Actually, one usually considers the SR for the operators $\Pi^{i\,OPE}_1(q^2)=32\pi^4\Pi^{i\,OPE}(q^2)$.
The factor $32\pi^4$ is
introduced in order to deal with the values of the order of unity (in
GeV units). After the Borel transform ${\cal B}$ we write (\ref{3a}) as
\begin{equation}
{\cal B}\Pi_1^{q}(q^2)=\sum_{n=0}A'_n(M^2); \quad {\cal B}\Pi_1^{I}(q^2)=\sum_{n=3} B'_n(M^2);
\quad A'_n(M^2)=32\pi^4{\cal B}A_n(q^2).
\label{3b}
\end{equation}
$$ B'_n(M^2)=32\pi^4{\cal B}B_n(q^2).$$
Here we present the most important terms
\begin{equation}
A'_0(M^2)=M^6; \quad A'_4(M^2)=\frac{bM^2}{4}; \quad
A'_6=\frac43 a^2;\quad
B'_3(M^2)=2aM^4.
\label{22a}
\end{equation}
The Borel transformed SR (\ref{3}) can be written now as
\begin{equation}
{\cal B}\Pi_1^i(M^2)={\cal F}_p^{i}(M^2)+{\cal F}_c^{i}(M^2),
\label{3g}
\end{equation}
where the two terms on the RHS are the contribution of the pole and that of the continuum:
\begin{equation}
{\cal F}_p^{i}(M^2)=\xi_i\lambda^2e^{-m^2/M^2};
\quad {\cal F}_c^{i}(M^2)=\int_{W^2}^{\infty}dk^2e^{-k^2/M^2}\Delta[{\cal B}\Pi_1(k^2)].
\label{3h}
\end{equation}

The conventional form of the SR is
\begin{equation}
{\cal L}^q(M^2, W^2)=R^q(M^2),
\label{4}
\end{equation}
and
\begin{equation}
{\cal L}^I(M^2, W^2)=R^I(M^2).
\label{4a}
\end{equation}
Here ${\cal L}^i$ and $R^i$ are the Borel transforms of the LHS and of the RHS of Eqs.~(\ref{3}), correspondingly
\begin{equation}
R^q(M^2)=\lambda^2e^{-m^2/M^2}; \quad R^I(M^2)=m\lambda^2e^{-m^2/M^2},
\label{5}
\end{equation}
with $\lambda^2=32\pi^4\lambda_N^2$. The contribution of the continuum is moved to the LHS of
Eqs.~(\ref{4}, \ref{4a}) which can be written as
\begin{equation}
{\cal L}^q=\sum_{n=0} \tilde A_n(M^2, W^2); \quad {\cal L}^I=\sum_{n=3}\tilde B_n(M^2, W^2),
\label{5a}
\end{equation}
-see Eq.~(\ref{3b}). Where

\begin{equation}
\tilde A_0=\frac{M^6E_2(\gamma)}{L(M^2)};\quad
\tilde A_4=\frac{bM^2E_0
(\gamma)}{4L(M^2)};
\label{22}
\end{equation}
$$\tilde A_6=\frac43 a^2L{(M^2)}; \quad
\tilde B_3=2aM^4E_1(\gamma); \quad \gamma=\frac{W^2}{M^2},$$
with
\begin{equation}
E_0(\gamma)=1-e^{-\gamma}, \quad E_1(\gamma)=1-(1+\gamma)e^{-\gamma}, \quad E_2(\gamma)=1-(1+\gamma+\gamma^2/2)e^{-\gamma}.
\label{23} \end{equation}

The factor
\begin{equation}
L(M^2)=\Big(\frac{\ln M^2/\Lambda^2}{\ln \mu^2/\Lambda^2}\Big)^{4/9}
\label{3100} \end{equation}
includes the most important radiative corrections of the order $\alpha_s\ln{(Q^2)}$~\footnote{Here the one loop running $\alpha_s$ with three light quark was used.}. These contributions were summed to all orders of $(\alpha_s\ln{Q^2})^n$.
In Eq.~(\ref{3100}) $\Lambda=\Lambda_{QCD}$ is the QCD scale, while $\mu$ is the
normalization point, the standard choice is $\mu=0.5\,$GeV.\\

\section{Polarization operator in instanton medium}
The nucleon polarization operator is driven by the "three quark" Green function which describes the propagation of three quarks from the point $x$ to the point $0$.
Recall that the quarks propagate not in the empty vacuum. The vacuum is filled by the instantons which modify
the quark Green functions.

\subsection{One instanton approximation}
Let us assume that all the vacuum expectation values of different OPE operators are caused by the instantons. Moreover the separation,  $R$, between the instantons is relatively large. After the averaging over the instanton orientation there is an effective repulsion between the instantons; a reasonable estimate is $R\sim 1$ fm~\cite{5,7,8,6} which is much larger than the inverse Borel mass, $1/M\sim 0.2$ fm in the duality interval used for sum rule solution.
Starting with the large negative $q^2$ on the LHS of sum rules we deal with the case when in the domain of interest, that is in the domain occupied by the polarization operator (\ref{6}), there is only one instanton. Note that here we consider the usual approximation where the modification of the quark propagators is caused just by the zero fermion mode in the instanton field, neglecting the possible contribution of higher modes; the quark propagator is written as a sum of the free quark (pertubative QCD) propagator and the zero mode contribution.

 In such a situation the instanton can
modify the propagator of only one quark (the d-quark in the proton) in the polarization operator formed by the Ioffe current (\ref{8}).
There is no 4-quark $\langle \bar u\bar uuu\rangle$ contribution since only one $u$ quark can be placed in the zero mode of the instanton. On the other hand the two quark $ud$,
 instanton induced contribution (that is the contribution of the instanton
induced $\langle \bar u\bar dud\rangle$ condensate)
 vanishes for the current (\ref{8}).
 Therefore in (\ref{3a}) the only non-vanishing terms are $A_0$ (Fig.~1) - the pure
 perturbative contribution and $B_3$ - which accounts for the $\langle |\bar q(0)q(x)|\rangle$
condensate.  In such a model there are no four and six ($\langle \bar q\bar q qq\rangle$ and
$\langle \bar q\bar q\bar q qqq\rangle$) quark condensate contributions. Correspondingly, the term $A_4$ which accounts for the quark-quark correlation in the "vacuum" gluon field absents as well. One can say that (in our model) these correlations nullify the contribution of the $\langle \bar q\bar q qq\rangle$ and $\langle \bar q\bar q\bar q qqq\rangle$ condensates in the case of the Ioffe current (\ref{8}).

However now we account for the $x$-dependence (non-locality) of the
$\langle |\bar qq|\rangle$ condensate.
This can be done in terms of the effective (dynamical) quark mass, $m(p)$. It was shown that the quark propagation in instanton medium  may be described by the Green function, which in the momentum representation takes the form~\cite{7,8}
\begin{equation}
S(p) = \frac{\hat p+im(p)}{p^2+m^2(p)} = S_I(p)+S_Q(p),
\label{31x}
\end{equation}
with $S_I$ and $S_Q$ corresponding to the structures proportional to
the unit matrix $I$ and to $\hat p$.

In our case, when we consider  a small space-domain where the quark interacts with one instanton only we write:
 \begin{equation}
S_Q(p)=\frac{\hat p}{p^2}; \quad S_I(p)=i\frac{m(p)}{p^2}\ .
\label{32}
\end{equation}
That is, following the previous discussion, we must include only the terms linear in $m(p)$. The momentum dependence of $m(p)$ reflects the $x$ dependence of the zero mode quark wave function while the normalization, $m(0)$, is driven by the value of local $\langle|\bar q(0)q(0)|\rangle$ condensate (see (\ref{54}) in the next section).
Since all propagators are diagonal in color, we shall present the contributions summed over the color
 indices.
Here $S_I$ is the zero-mode contribution. The sum of all the nonzero-mode contributions  is approximated thus by the free propagator of the massless quark $S_Q(p)$.

\subsection{Contributions to polarization operator}

The leading contribution  $A_0$ to the $\hat Q$ structure
remains unchanged . However, the leading contribution $B_3$ to the $I$ structure is now~\footnote{The results are obtained in \cite{DS}.}
\begin{equation}
\Pi^I(Q^2)=12\int\frac{d^4p}{(2\pi)^4}\gamma_{\mu}\frac{m(p)}{p^2}\gamma_{\nu}T_{\mu\nu}(Q-p),
\label{48}
\end{equation}
while
\begin{equation}
T_{\mu\nu}(Q-p)=\int d^4x e^{-i(Q-p,x)}Tr[t_{\mu\nu}(x)],
\label{35}
\end{equation}
with
\begin{equation}
t_{\mu\nu}(x)=\gamma_{\mu}G_0(x)\gamma_{\nu}G_0(x).
\label{36}
\end{equation}
Here
\begin{equation}
G_0(x)=-\frac{1}{2\pi^2}\frac{\hat x}{x^4}
\label{37}
\end{equation}
is the Fourier transform of the propagator $S_Q$ determined by Eq.~(\ref{32}).

To obtain results in analytical form we parameterize the dynamical quark mass caused by the small size instantons
\begin{equation}
m(p)=\frac{{\cal A}}{(p^2+\eta^2)^3},
\label{53}
\end{equation}
with ${\cal A}$ and $\eta$ the adjusting parameters. The power of denominator insures the proper behavior
$m(p) \sim p^{-6}$ at $p \rightarrow \infty$ \cite{7}.

In terms of the parametrization (\ref{53})
the quark condensate created by the field of instantons  can be written as
\begin{equation}
\langle  0|\bar q(0)q(0)|0\rangle_s
=-4N_c\int\frac{d^4p}{(2\pi)^4}\frac {m(p)}{p^2}=\frac{3{\cal A}}{2\eta^4}\ .
\label{54}
\end{equation}
(Here we kept the Minkowsky metric for the quark operators in the vacuum expectation values.)

Calculating the tensor $T_{\mu\nu}$ we present
\begin{equation}
 \Pi^I(Q^2) =\frac{3}{\pi^2}\int \frac{d^4p}{(2\pi)^4} \frac{{\cal A}}{p^2(p^2+\eta^2)^3}(Q-p)^2\ln{(Q-p)^2},
\label{49}
\end{equation}

Note that putting $p=0$ in the factor $(Q-p)^2\ln{(Q-p)^2}$ of the integrand on the RHS of Eq.~(\ref{49}) we would obtain
\begin{equation}
\Pi^I(Q^2)=X_s=\frac{3Q^2\ln{Q^2}}{8\pi^4}\int_0^{\infty} dp\,p\,m(p)=B_3(Q^2),
\label{50}
\end{equation}
with $B_3(Q^2)$ defined by Eq.~(\ref{3c}).
Thus we can view calculation of the contribution given by Eq.~(\ref{48}) as inclusion of non-localities in the scalar quark condensate.

Further details of calculations are presented in  \cite{DS}. We have
\begin{equation}
B'_3(M^2)=
2aM^4F(\beta); \quad F(\beta)=\frac{1}{3}\Big(\frac{2(1-e^{-\beta})}{\beta}+e^{-\beta}(1-\beta) +\beta^2{\cal E}(\beta)\Big).
\label{55}
\end{equation}
$$\beta=\eta^2/M^2.$$
Here
\begin{equation}
{\cal E}(\beta)=\int_{\beta}^{\infty}dt\frac{e^{-t}}{t}.
\label{56}
\end{equation}
Adding the contribution of continuum states we obtain \cite{DS}
\begin{equation}
\tilde B_3(M^2,W^2)= 2a\,M^4\,\Phi(M^2, W^2)\,;
\label{67}
\end{equation}
$$\Phi(M^2, W^2)=\frac{1}{3}\Big(\frac{2}{\beta}(1-e^{-\beta})+e^{-\beta}(1-\beta)-e^{-\gamma}(1-\beta+\gamma) +\beta^2({\cal E}(\beta)- {\cal E}(\gamma)) \Big)\,.$$
The functions $E_i (i=0,1,2)$ are determined by Eq.~(\ref{23}).

\subsection{Solutions}
The solutions of sum rules for the nucleon in conventional approach and in the instanton vacuum are presented in Tables 1,2. Here we choose the  values of instanton radius $\rho=0.33$ fm and separation $R\simeq 1$ fm which provide the quark condensate $a=0.55\,{\rm GeV}^3$.
We account for the anomalous dimension of the operators (factor (\ref{3100}) in (\ref{22})) and for the non-logarithmic $\alpha_s$ correction~\cite{a-cor} implemented as it described in~\cite{PRC80}~\footnote{Strictly speaking, in the case of the instanton zero mode contribution the $\alpha_s$ correction may be a little bit different, however we keep the same (old) expression to see just the difference caused by the non-locality of the quark condensate. We checked that the variation of the correction factor (\ref{b3}) between $1+\alpha_s/\pi$ and $1+2\alpha_s/\pi$ changes the value of nucleon mass $m$ less than $\pm 3\%$.}.

That is the terms $A_0,\ B_3$ and $A_6$ are multiplied by the factors  (see Eqs.~(18),(23) in \cite{PRC80})
\begin{equation}\label{b1}
\tilde A_0\rightarrow  \tilde A_0\cdot r_0;\quad r_0=1 + \frac{\alpha_s}{\pi}\,\frac{53}{12}.
\end{equation}

\begin{equation}
\label{b3}
\tilde B_3\rightarrow  \tilde B_3\cdot r_3; \quad r_3= 1+ \frac{\alpha_s}{\pi}\,\frac32.
\end{equation}

\begin{equation}\label{b2}
\tilde A_6\rightarrow  \frac43\,\frac{a_4(M^2)}{L(M^2)}\cdot r_6; \quad a_4(M^2)=a^2\,L(M_0^2);
\end{equation}
$$
r_6 = 1 - \frac{\alpha_s}{3\pi}\,\left[\frac52 + \ln\frac{W^2}{M_0^2} + {\cal E}(-W^2/M^2)
\right].
$$
Here ${\cal E}(x)=\sum_{n=1} x^n/(n\cdot n!)$.

We use the quark condensate $a=0.55\,{\rm GeV}^3$, $\Lambda_{QCD}=230$~MeV and the value of QCD coupling is found
 at the scale 1~GeV; that is the corresponding one loop (with three
light quarks) $\alpha_s=0.475$ in (\ref{b1},\,\ref{b2},\,\ref{b3}). In \cite{PRC80} it was assumed that the factorization of the $4$-quark condensate takes place at a certain momentum $k$: $k^2=M_0^2$, $M_0^2=1\,{\rm GeV}^2$.
The duality interval
\begin{equation}\label{M2}
 0.8 < M^2 < 1.4 \mbox{GeV}^2 .
\end{equation}
In the line $I$ of the Table 1 the nucleon parameters obtained in the full sum rules equations \cite{PRC80} are shown. The difference between presented computation and  \cite{PRC80} is in the value of $\Lambda_{QCD}$.  The full sum rules equations contain some more terms in comparison with Eqs.~(\ref{3c}), they are: the term $A_8$ which accounts for the $\langle|\bar q\bar q G_{\mu\nu}\sigma_{\mu\nu}qq|\rangle$  condensate; $B_7$ - containing  both quark and gluon condensates; $B_9$ - which is the 6q-condensate. The values of these condensates are assumed to be the same as in~\cite{PRC80}. In the line $II$ we present the results obtained when only the terms $A_0$ and $B_3$ with $\alpha_s$-corrections and anomalous dimensions are taken into account in calculations.

As it is seen from Table 1 accounting for the structure of instanton vacuum we get the result quite different from that obtained in the framework of the conventional approach, which is shown in line $I$.
 The difference is caused by two effects. First, it is the absence of the
four quark condensate contribution.
In the one instanton model with the  Ioffe current (\ref{8}) we have $A_6=0$;
the corresponding result where only the terms $A_0$ and $B_3$ are included is presented in line $II$. Next it is  the $x$-dependence (non-locality) of the $\langle 0|\bar q(0)q(x)|0\rangle$ condensate.   After the inclusion
   of non-locality in a simple one instanton model we actually loose the
solution. The
values of the residue $\lambda^2$, the continuum threshold, $W^2$ and of the nucleon mass $m$ become too small and the contribution of nucleon pole becomes negligible.\\

Finally we consider the situation where only a fraction, $w_s<1$, of the conventional $\langle |\bar qq|\rangle$ condensate is due to the small size instantons with $1/\rho\simeq 600$~MeV; the remaining part of the condensate is written in the first term of Eq.~(\ref{34a}) as  $(1-w_s)\langle |\bar q(0)q(0)|\rangle$. This remaining part is caused by some slowly varying color field or by  large
size instantons and can be included in our calculations in a "local" approximation, that is replacing $\langle|\bar q(0)q(x)|\rangle$ by
$\langle|\bar q(0)q(0)|\rangle$.

In other words now we will keep  the  contributions $A_0$, $B_3$ and $A_6$ in the sum rules for the nucleon (\ref{3a}). The $\Pi^I$ component takes the form
 \begin{equation}
\Pi^I(Q^2)=2a(1-w_s)Q^2\ln{Q^2}+w_sX_s\ ,
\label{34a}
\end{equation}
where $X_s$ is given by the previous expression (\ref{50}).

The four quark condensate contribution with one quark placed in the instanton zero mode can be written as
\begin{equation}
A_6=\frac{4a(1-w_s)w_s}{\pi^2}\int \frac{d^4p}{(2\pi)^4}\,\frac{m(p)}{p^2}\,
\frac{\hat Q-\hat p}{(Q-p)^2}\ ,
\label{63}
\end{equation}
leading to the additional factor in expression for $\tilde A_6$ (\ref{b2}) \cite{DS}. Thus
\begin{equation}
\tilde A_6 \to \tilde A_6\,(1-w_s)\,\left((1-w_s)+2w_s\frac{E_0(\beta)}\beta \right)\,.
\label{63c}
\end{equation}
This corresponds to the case when the remaining $(1-w_s)\langle |\bar qq|\rangle$ condensate is produced by some completely different slowly varying colour configuration and there is no repulsion between this long-wave configuration and  the small size instanton. Here the first term in (\ref{63c}) corresponds to the case when both quarks are coming from this slowly varying field while the second term accounts for one quark in the instanton zero mode. Otherwise we still put $A_6=0$ {\em if} there is the repulsion between these two sources of quark condensate. Say, when the remaining part is caused by the larger size instantons which is due to instanton-instanton effective repulsion can not occupy the same space domain  we  put $A_6=0$~\footnote{In the last case we have to say (assume) that in the domain occupied by the polarization operator
(\ref{6}) there is only {\em one}, either large or small size instanton.}.
We consider both possibilities.
The set of diagrams included in the SR is shown in Fig.~1.\\

To demonstrate the size of a possible effect we take a 50\% - 50\% mixture of a small ($1/\rho=600$~MeV) and a large ($1/\rho\gg M_{\mbox{Borel}}$) size configurations.
The results shown in Table 2 correspond to solution of SR equations with the conventional total quark condensate $a=0.55\,{\rm GeV}^3$
(the quark mass at $p=0$ is $m(0)=0.292$~GeV) but the contribution of a small size configurations are taken with the weight $w_s=0.5$.

Assuming that $A_6=0$ due to repulsion, we get a reasonable solution (lines $I$) with
$m=1063\,{\rm MeV}$, $\lambda^2=1.40\,{\rm GeV}^6$, $W^2=1.97\,{\rm GeV}^2$ and very good "duality" manifested by $\chi_N^2=0.002$. The ratios of the RHS/LHS inside the duality interval (\ref{M2})
 are shown in Fig.~2.
Omitting the $\alpha_s$-corrections we obtain too large value of mass $m=1268$~MeV.\\

If we neglect the repulsion effect and include the four quark $A_6$ term then we loose the physical solution. Both - with and without the $\alpha_s$-corrections, the contribution of the nucleon pole term becomes too small (see the lines $II$). However if we fix the value of $m=940$ MeV (lines $III)$ than the pole contribution becomes larger and the quality of 'duality' between the LHS and the RHS of sum rules equations is not too bad, corresponding to about 10\% error. On the other hand such a solution does not provide the global minimum of $\chi^2$.

The dependence of the parameters $m,\ \lambda^2$ and $W^2$ on the position of duality interval is shown in Table 3. It is seen that in spite of the fact that for a larger Borel mass  we get a much better $\chi^2$, the relative nucleon pole contribution strongly decreases with $M^2$. On the other hand the value of $m$ depends on the duality interval weakly.  Therefore the conventional $0.8 < M^2 < 1.4$~GeV$^2$ interval (\ref{M2}) looks as an optimal choice.

\section{Comment about the previous attempt to account for the instantons}
The instanton contribution to the QCD sum rules for the nucleon was discussed long ago in~\cite{12,13,14}.    Authors were interested in description of the quark correlations in the polarization operator due to interaction of two quarks with the same instanton. In other words the authors  considered the instanton correction to the Green function of {\em two} quarks but neglect the instanton effect in the propagation of one quark where just the "mean field" scalar condensate $\langle 0|\bar q(0)q(0)|0\rangle$ was included.

Moreover in the case of Ioffe current (\ref{8}) the term considered in these papers \cite{12,13,14} have the form
\begin{equation}\label{KD}
\langle |\bar qq|\rangle\cdot |\mbox{wave function of two quarks in instanton zero mode}|^2.
\end{equation}
Simultaneously the authors choose such  instanton parameters that the whole value of the  condensate $\langle |\bar qq|\rangle$
is produced by the same instanton. As far as we account for the instanton-instanton repulsion it means that we are looking for
two $u$-quarks (in a proton case) in the same zero mode, that is impossible. The only possibility is to say that the $\langle |\bar qq|\rangle$
factor in (\ref{KD}) is of another (not the same instanton) origin. This means that this factor should be smaller
(most probably much smaller) than the conventional value of $\langle |\bar qq|\rangle$ condensate; that
is the effect should be much smaller (at least it should be suppressed by the factor $w_s(1-w_s)< 1/4$)
than the estimation presented in these papers.

\section{Conclusion}
We have demonstrated that the non-trivial structure of quark condensate caused
by the quark-quark correlations in the instanton colour field strongly affected the solution of QCD sum rules for the nucleon. In a simplified model of instanton vacuum where the size of instanton is fixed ($1/\rho=600$~MeV) and  only one instanton can be placed into the domain occupied by the nucleon polarization operator (\ref{6}) we loose the solution at all. However assuming that only one half of the $\langle|\bar q(0)q(x)|\rangle$ condensate has the non-locality induced by the fermion zero mode of the small size $\rho=1/(600\mbox{MeV})$ instantons   we obtain a reasonable solution with $m=1063$~MeV (shown in line $I$ of the Table~2). Thus the QCD sum rules still can be used to evaluate the properties of the hadrons but one has to take care about more complicated structure and the $x$-dependence of a more realistic condensates.

However now we have to account for the $x$-dependence of condensates explicitly. The inverse size of the instanton  is comparable to the typical Borel mass  used in the QCD sum rules. Therefore we have no reasons to expect a good convergency of the OPE  and it is not sufficient to include the non-locality just in terms of the first (or few) derivative(s) corresponding to the Taylor expansion at $x^2=0$.

\vspace{1cm}

{\bf Acknowledgements}\\
We thank A.E. Dorokhov, N. I. Kochelev and especially V. Yu. Petrov for illuminating discussions.
We also acknowledge the partial support by the RFBR grant 12-02-00158.

{}

\newpage

\begin{table} 
\begin{center}
\begin{tabular}
{|c|c|c|c|c|c|} \hline
version &$m$ &$\lambda^2$&$W^2$&$\chi_N^2$ & \\
 & MeV & GeV$^6$ & GeV$^2$& &\\
\hline
$I$ &928&2.36&2.13&0.02&  \\
\hline
$II$  &1271&4.35&3.09&0.02 & \\ 
\hline
\hline
$III$ & 199&0.04&0.63&0.05&non-physical\\
\hline
$III(\mbox{no}\, \alpha_s)$&526&0.041&0.756&0.012& non-physical\\
\hline
$III$($m$\ -\ fixed)&940&0.52&1.46&7.70& non-physical\\
\hline
\end{tabular} \end{center}
\caption{Solutions of the sum rules equations.
In the line $I$  calculations are made by the  Ioffe equations (see text);
 the line $II$: parameters are obtained with $A_0$ and $B_3$ terms  only;
the line $III$: calculations are made using the contribution of the single instanton by equation (\ref{67})
and  $\alpha_s$-corrections are taken into account;
$III(\mbox{no}\, \alpha_s)$ - the same as $III$ but without the  $\alpha_s$-corrections
 (\ref{b1} - \ref{b2});
$III$($m$-fixed) - the same as $III$ but mass is taken equal to $m=0.94$~GeV and fixed. $\chi^2$ values are calculated assuming the 10\% error bars, $\chi_N^2=\chi^2/N$, $N$ is the number of fitting point.
}
\end{table}

\begin{table}
\begin{center}
\begin{tabular}
{|c|c|c|c|c|c|} \hline
version &$m$ &$\lambda^2$&$W^2$&$\chi_N^2$ &\\
 & MeV & GeV$^6$ & GeV$^2$& &\\
\hline
$I$  &{\bf 1063}& {\bf 1.40}& {\bf 1.97} & {\bf 0.002} &\\
\hline
$I(\mbox{no}\, \alpha_s)$ &1268&2.21&3.02&0.02 &\\ 
\hline
$II$ &574&0.55&1.07&0.09& \mbox{non-physical} \\  
\hline
$II(\mbox{no}\, \alpha_s)$  &640&0.60&1.34 &0.11&\mbox{non-physical} \\ 
\hline
$III$ & 940&1.46&1.89&1.22&$m$\ -\ fixed\\
\hline
$III(\mbox{no}\, \alpha_s)$&940&1.36&2.36&0.46&$m$\ -\ fixed\\
\hline
\end{tabular} \end{center}
\caption{Solutions of the sum rules equations  assuming that the non-locality of  instanton induced quark condensate is included with the weight $w_s=0.5$.
 In the lines $I$ and $I(\mbox{no}\, \alpha_s)$  calculations are made by the SR equations with the following terms on the LHS: $\tilde A_0$ (\ref{22}) and $\tilde B_3$ (\ref{67}). The $\alpha_s$-corrections are included in results of $I$ and not included in $I(\mbox{no}\, \alpha_s)$.
Results in the lines $II,\ III$ are obtained with $\tilde A_0$, $\tilde B_3$ and
$\tilde A_6$ given by Eqs.~(\ref{22}), (\ref{67}), (\ref{63c}) in the SR left hand sides. The $\alpha_s$-corrections are included in line $II$ and not included in $II(\mbox{no}\, \alpha_s)$. In lines $III$  we present the same results as in $II$ but for the fixed $m=0.94$~GeV.}
\end{table}

\begin{table}
\caption{Solutions of the sum rules equations in various intervals of the values of the Borel mass.}
\begin{center}
\begin{tabular}{|c|c|c|c|c|} \hline
$M^2,\,{\rm GeV}^2$ & $m,\,{\rm MeV}$ & $\lambda^2,\,{\rm GeV}^6$ & $W^2,\,{\rm GeV}^2$ & $\chi_N^2$\\
\hline

0.8-1.4&1063&1.40&1.97&1.6(-3) \\

1.0-1.6&1066&1.37&1.94 &2.3(-4)\\

1.4-2.0&1069&1.36&1.93&1.0(-5)\\

\hline
\end{tabular} \end{center}
\end{table}


\clearpage

\newpage

\section*{Figure captions}

\noindent
Fig.~1. Diagrams for the nucleon polarization operator in the presence of the
instanton; ($a$) - free quark propagators, ($b,c$) - one quark is in the zero
fermion mode, ($d,e$) - one quark reverse the helicity in the zero mode of
the small instanton while another quark change its helicity interacting
with the long wave colour field (large instanton or another long wave
field).

\noindent
Fig.~2  The ratios of the RHS/LHS inside the duality interval (\ref{M2}) for the parameters on the line $I$ in Table 2.
The solid and dashed lines are for the $\hat q$ and $I$ parts of the sum rules, correspondingly.

\newpage

\begin{figure}
\centerline{\epsfig{file=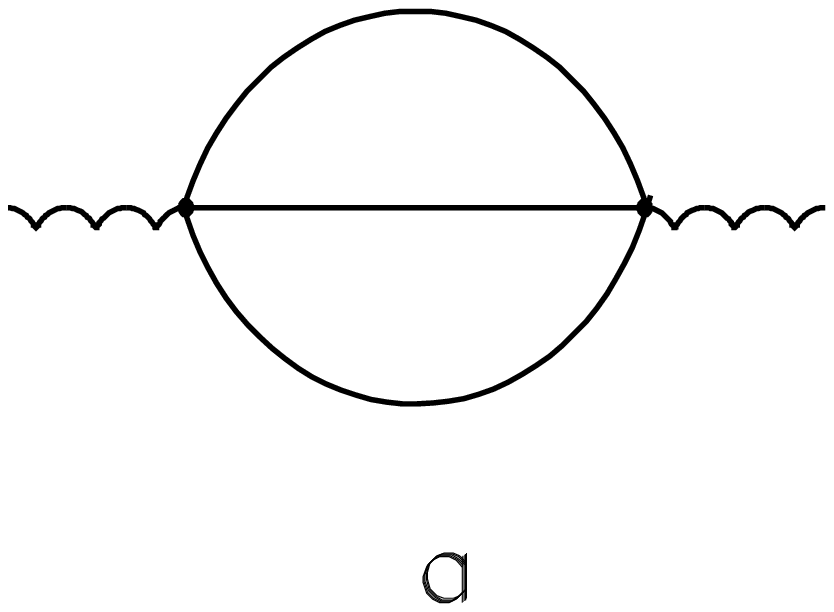,width=5cm}}
\centerline{\epsfig{file=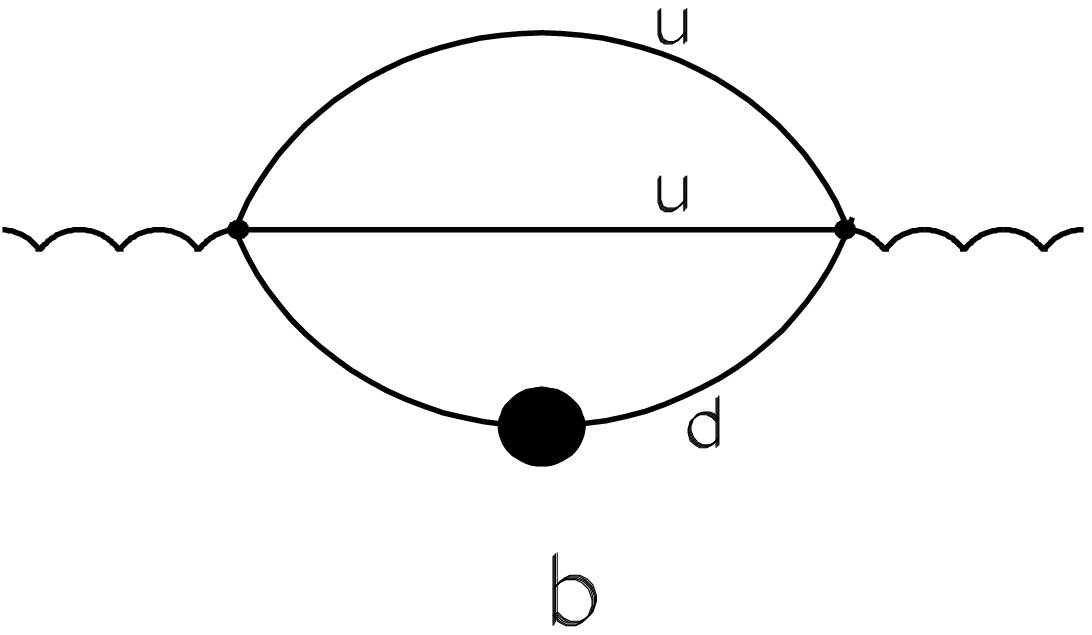,width=5cm} \hspace{0.5cm}\epsfig{file=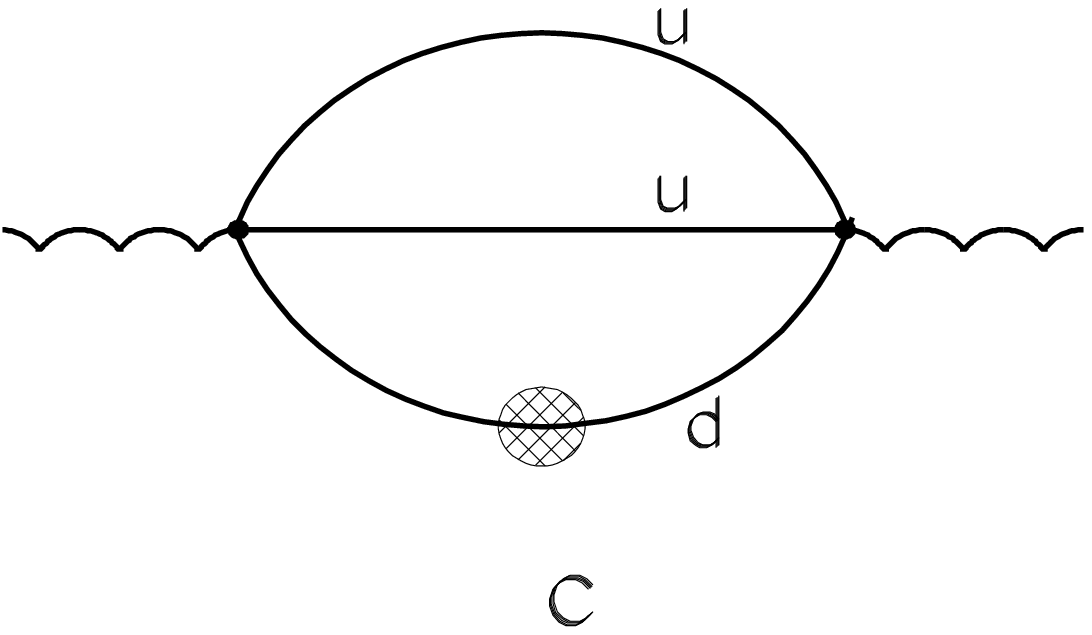,width=5cm}}
\centerline{\epsfig{file=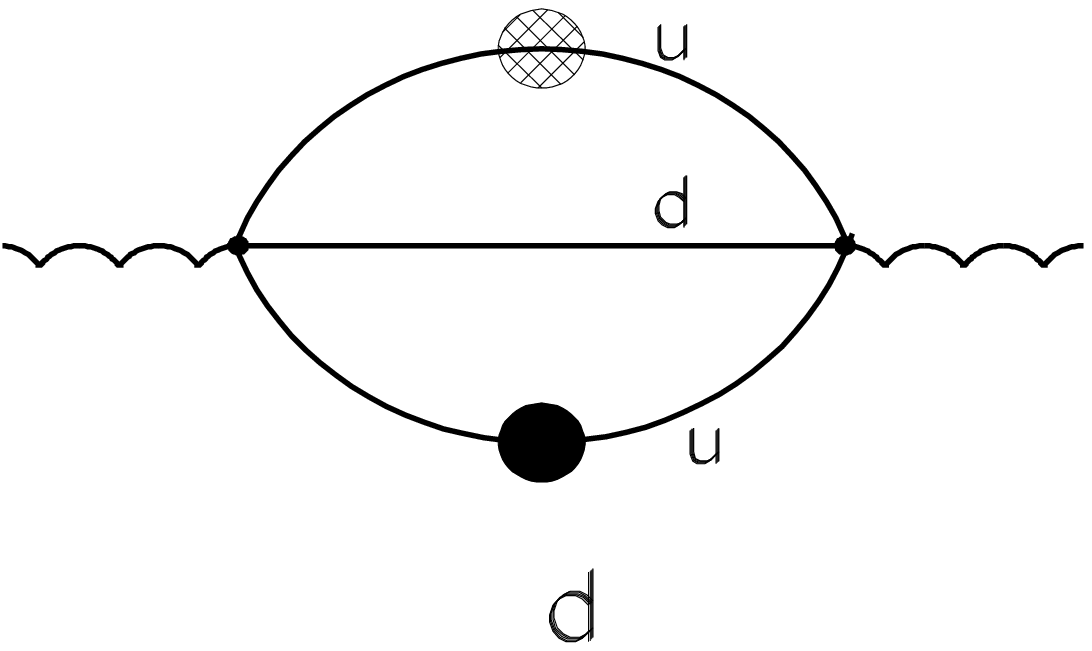,width=5cm} \hspace{0.5cm}\epsfig{file=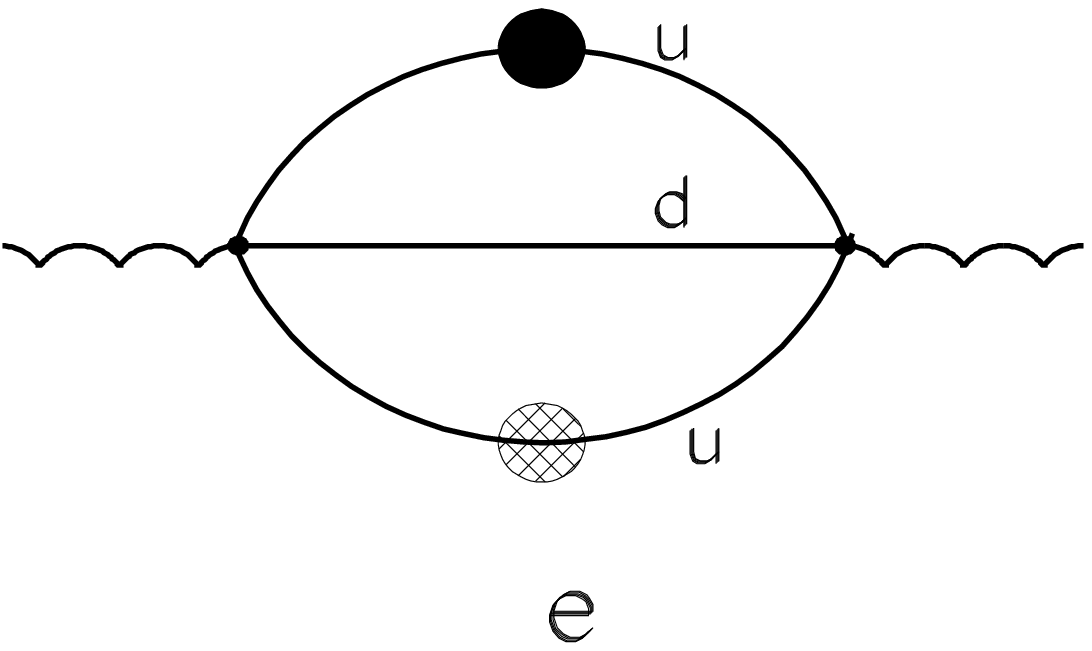,width=5cm}}
\caption{}
\end{figure}

\begin{figure}
\centerline{\epsfig{file=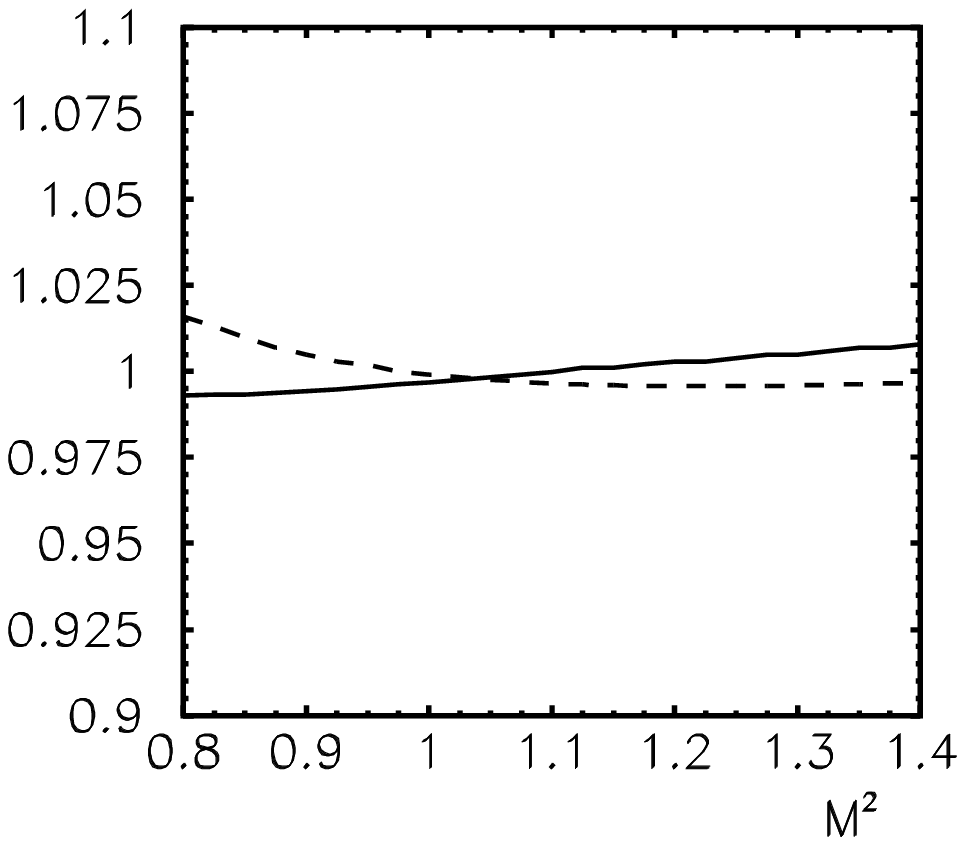,width=6cm}}
\caption{}
\end{figure}


\newpage

\begin{thebibliography}{}

\bibitem{1}M.A. Shifman, A.I. Vainshtein and V.I.~Zakharov, Nucl.
Phys. ~{\bf B147}, 385; 448; 519 (1979).

\bibitem{2} B.L. Ioffe, Nucl. Phys. ~{\bf B 188}, 317 (1981); {\bf B 191}, 591(E) (1981).
\bibitem{3} B.L. Ioffe, L.N. Lipatov and V.S.~Fadin, {\em Quantum
Chromodynamics} (Campridge Univ. Press, 2010).
\bibitem{9} B.L. Ioffe, Z. Phys. C {\bf18}, 67 (1983).
\bibitem{BI} V.M. Belyaev and B.L.~Ioffe, Sov. Phys. JETP {\bf 56}, 493 (1982).
\bibitem{Si} E.V.~Shuryak, Nucl. Phys. {\bf B328}, 85 (1989).
\bibitem{MR} S.V. Mikhailov and A.V. Radyushkin, JETP Lett. {\bf 43} 712 (1986);
 Phys. Rev. {\bf D45}, 1754 (1992).
\bibitem{5}E. V. Shuryak, {\em The QCD Vacuum, Hadrons and the Superdense Matter}, World Scientific Pub. Co, Singapore, 1988.\\
T. Sch\"afer and E.V.~Shuryak, hep-ph/9610451.
\bibitem{7}  D. I. Dyakonov and V. Yu. Petrov, Sov. Phys. ZhETP, {\bf 89}, 361 (1985).
\bibitem{8}  D. I. Dyakonov and V. Yu. Petrov, Nucl. Phys. B {\bf 272}, 457 (1986).
\bibitem{6} A. Ringwald, F. Schrempp, Phys. Lett. {\bf B459}, 249 (1999).
\bibitem{DS} E. G. Drukarev, V. A. Sadovnikova, eprint arxiv.org: 1407.2749 [hep-ph].

\bibitem{a-cor} M.~Jamin, Z. Phys. {\bf C37}, 635 (1988);\\
Y.~Chung, H.G.~Dosch, M.~Kremer and D.~Schall, Z. Phys. {\bf C25}, 151 (1984);\\
A.A.~Ovchinikov, A.A.~Pivovarov and L.R.~Surguladze, Int. J. Mod. Phys. {\bf A6}, 2025 (1991).
\bibitem{PRC80} E. G. Drukarev, M. G. Ryskin, V. A. Sadovnikova, Phys. Rev. {\bf C80}, 045208 (2009).
\bibitem{12} A. E. Dorokhov, N. I. Kochelev, Z. Phys. C~{\bf 46}, 281
(1990).
\bibitem{13} H. Forkel, M. K. Banerjee, Phys. Rev. Lett. {\bf71}, 484
(1993).
\bibitem{14} Hee-Jung Lee, N. I. Kochelev, V. Vento, Phys. Lett. B~{\bf 610}, 50
(2005).

\end{thebibliography}
\end{document}